# Simple alternative model of the dual nature of light and its *Gedanken* experiment

François Hénault
UMR 6525 H. Fizeau, Université de Nice-Sophia Antipolis
Centre National de la Recherche Scientifique, Observatoire de la Côte d'Azur
Parc Valrose, 06108 Nice – France

**ABSTRACT**

In this paper is presented a simple alternative model of the dual nature of light, based on the deliberate inversion of the original statement from P. A. M. Dirac: "Each photon interferes only with itself. Interference between different photons never occurs." Such an inversion implies that photons and light quanta are considered as different classes of objects, but stays apparently compatible with results reported from different recent experiments. A *Gedanken* experiment having the potential capacity to test the proposed model in single photon regime is described, and its possible outcomes are discussed. The proposed setup could also be utilized to assess the modern interpretation of the principle of complementarity.

## 1  INTRODUCTION

Questions about the dual nature of light have been puzzling most physicists since more than one century, when M. Planck and A. Einstein were conducted to introduce the concept of "quantum of light", later called photon. From this moment, light was considered either as a wave (the classical theory) or a stream of particles exchanging their energy and momentum with electrons or atoms (quantum electrodynamics). Moreover, according to N. Bohr's principle of complementarity [1], it should not be possible to conceive a physical experiment demonstrating both particle and wave behaviours of light at the same time. The purpose of this paper is twofold: first, a simple alternative representation of the nature of light is proposed, based on the assumption that photons and light quanta are different classes of objects. Second, a thought (or *Gedanken*) experiment is described, having the potential capacity to assess the model and secondarily to test the modern interpretation of the principle of complementarity.

## 2  THE MODEL

It is nowadays generally admitted that only quantum electrodynamics theory can provide a satisfactory explanation for the nature of light [2], either corpuscular or wave-like. Some scientists, however, remain in favour of the semi-classical model, associating a continuous electrical field with quantized absorption and emission processes [3]. In this respect the herein proposed model can be considered as a compromise between both theories, preserving the quantified nature of the light and classical wave comprehension. The model is based on the following hypotheses:

1) One quantum of light carries the smallest possible fraction of luminous energy and cannot interfere with itself.

2) Light quanta can only be divided spatially, e.g. by multiple apertures. When encountering a dielectric interface, they are either reflected or transmitted, but not both.

3) Light quanta are of ondulatory nature in the sense that they are carrying a phase related to their geometrical trajectories and the past events (e.g. aperture divisions or reflections and transmissions on dielectric interfaces).

4) All light quanta emitted by the same electronic transition of an atom (or different atoms) can interfere together. This interference process is governed by the established rules of spatial and temporal coherence (either classical or quantum).

5) When detected by a measurement apparatus, the complex amplitudes of all incoming light quanta are added coherently, and the resulting energy (expressed in terms of the number of photons) is equal to the square modulus of the whole complex amplitude. Hence the number of light quanta is not necessarily equal to the total number of photons: photons and light quanta are two different classes of objects.

The second and third hypotheses are necessary to take into account the results of Grangier, Roger and Aspect experiments in the single photon regime [4]: utilizing a two-photon radiative cascade, they observed no counting coincidences at both sides of a beamsplitter (hence photon is a particle), but when a Mach-Zehnder interferometer was constructed around the same beamsplitter, the two output ports of the interferometer revealed the typical phase opposition foreseen by classical interference theory. However the most unexpected and deranging assumptions are probably the first and fourth ones, sounding as a literal inversion of P. A. M. Dirac's famous hypothesis about photons only interfering with themselves, as written in his reference textbook [5]:

> "Suppose we have a beam of light consisting of a large number of photons split up into two components of equal intensity. On the assumption that the intensity of a beam is connected with the probable number of photons in it, we should have half the total number of photons going into each component. If the two components are now made to interfere, we should require a photon in one component to be able to interfere with one in the other. Sometimes these two photons would have to annihilate one another and other times they would have to produce four photons. This would contradict the conservation of energy. The new theory, which connects the wave function with probabilities for one photon, gets over the difficulty by making each photon go partly into each of the two components. Each photon then interferes only with itself. Interference between two different photons never occurs."

The final hypothesis is clearly derived from an energy conservation constraint where photons and light quanta are considered as equivalent particles. But one might also imagine that different light quanta can interfere together without contradicting the conservation of energy that should only be applicable to photons. In other words, this final assertion could be reversed as follows: "Each light quantum interferes only with another one. Interference of a single light quantum never occurs." Furthermore, it has already been pointed out that our modern interpretation of Dirac's sentence should not be understood too restrictively [6], since interference between separate laser or atoms sources are now admitted and commonly realized (see for example Refs. [7-9]). Also, inverting Dirac's famous statement stays compatible with some recent variants of the historical Young's double slit experiment, where the following results have been reported:

- The interference pattern remains unaffected when the slits are alternatively masked in phase opposition [10].

- It is again unaffected when the optical beams emitted from the source toward each individual slit are carefully baffled [11].

- Finally, it can survive to frequency up-conversion from infrared to visible spectral domains, operated by two or more separated nonlinear optical frequency converters [12].

But perhaps a more spectacular indication resides in the astronomical pictures that have been routinely produced by the Hubble Space Telescope since twenty years. Two examples of them are reproduced in Figure 1, showing diffraction effects generated by the telescope spider legs on bright, foreground stars, while fainter, background stars or galaxies are not affected with diffraction phenomena for identical integration times (it should be noticed that laboratory experiments using lasers led to similar observations, see chapter 7 of Ref. [3]). This suggests that many light quanta could interfere together, and that this interference phenomenon would fade away as less and less of them are present: here again Dirac's hypothesis has been inverted.

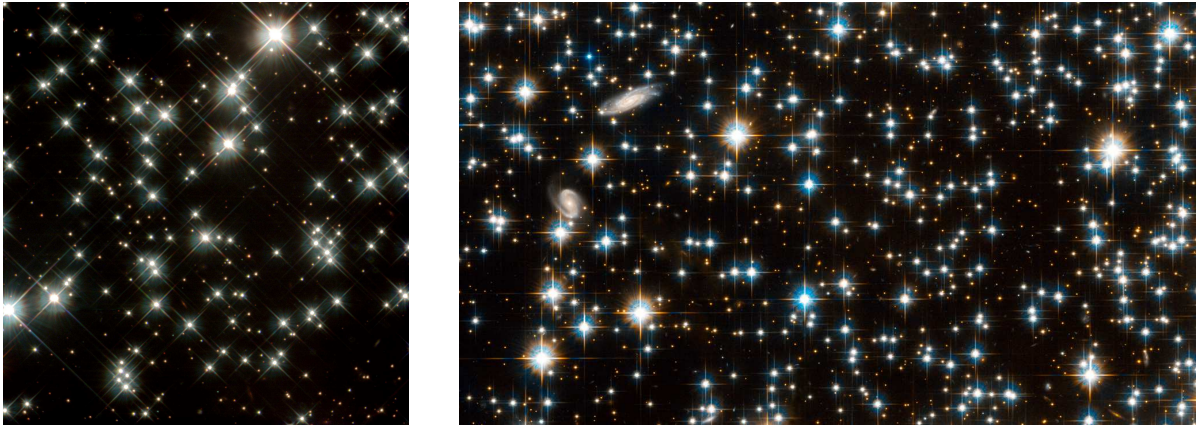

Figure 1: Pictures of globular star cluster M4 (left) and NGC 6791 (right) from the Hubble Space Telescope. Credits NASA/H. Richer (University of British Columbia) and NASA/ESA/Digitized Sky Survey/L. Bedin (STScI).

## 3   *GEDANKEN* EXPERIMENT

Although a model based upon hypotheses n° 1-5 in section 2 may look simplistic and naïve, it might be possible to define a thought, or *Gedanken* experiment in order to verify or refute it. For that purpose the latter should incorporate one source of single-photons, using for instance the technique of spontaneous parametric down-conversion, and also answer to two specific requirements, namely the division of the input beam by means of a dielectric plate in order to test assumptions n°2 and 3, and a multi-axial light recombination as in Young's double slit experiment to assess assumptions n°1 and 4. The proposed setup, somewhat inspired from ref. [13] is therefore depicted in Figure 2. The entrance light quanta are first directed towards an air glass interface where they can be reflected under a high incidence angle *i*, or transmitted with comparable probabilities. The glass plate, which plays the role of the beamsplitter, is actually a small prism whose angle is defined so as to reject parasitic internal reflections out of the interferometer. The reflected and transmitted beams are further recombined multi-axially by a concave mirror M focusing them near its focal point F (note that this monolithic mirror could be replaced by two separate flat mirrors). One of the beams (here the transmitted one) is slightly misaligned with respect to the other by means of the tip-tilt mirror M'. Finally, an optional reflective delay line can be added into the apparatus in order to equalize or modulate the optical path difference. The experiment is realizable for any state of polarization of the entrance light by simple readjustment of the angle *i* in order to equalize the amplitudes of both input beams.

One peculiarity of the proposed setup probably resides in its intrinsic asymmetry, because it starts as a Mach-Zehnder interferometer and ends up in a Young's double slit combining scheme, although most of the previously reported experiments are based on fully symmetric arrangements (either based on classical Mach-Zehnder or Young interferometers). Also, all photon-detecting devices can be installed in the same observation plane (i.e. the focal plane of mirror M).

The measurement apparatus itself essentially consists of four optical fibers located at accurate positions with respect to the expected intensity distributions when light is considered as either a particle or a wave (see Figure 3). The optical fibers are feeding two couples of photo-detectors denoted (P1,P2) and (W1,W2) where letters P and W respectively stand for "particle" and "wave" as in ref. [14]. The correlations and anti-correlations between detectors P1 and P2 are measured by a coincidence circuit as in the first step of the Grangier, Roger and Aspect experiment [4]. P1 and P2 are expected to provide a certain amount of "which-way" information that we note P = |P1-P2|. On the other hand, detectors W1 and W2 fed by the central fibers respectively located at two adjacent maxima and minima of the expected interferogram are measuring the fringe visibility W = |W1-W2| (in practice, this interference counter may only "click" if W is superior to a certain threshold). The principle of the experiment suggests that we could in principle obtain a simultaneous measurement of both P and W – here the original principle of complementarity would not be respected [1], but it has been shown in later publications that it can be re-expressed in a more general way as $P^2 + W^2 \leq 1$ [14-15].

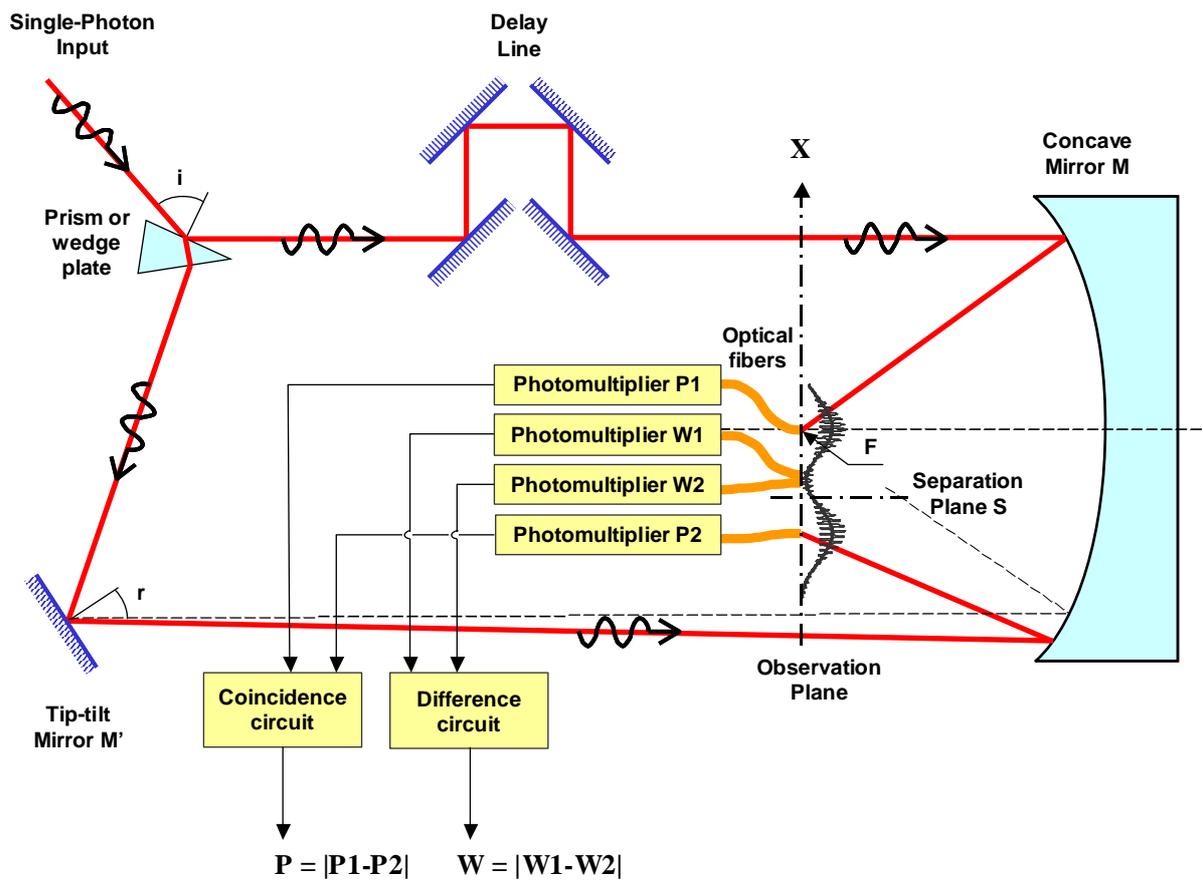

Figure 2: Principle of the proposed *Gedanken* experiment.

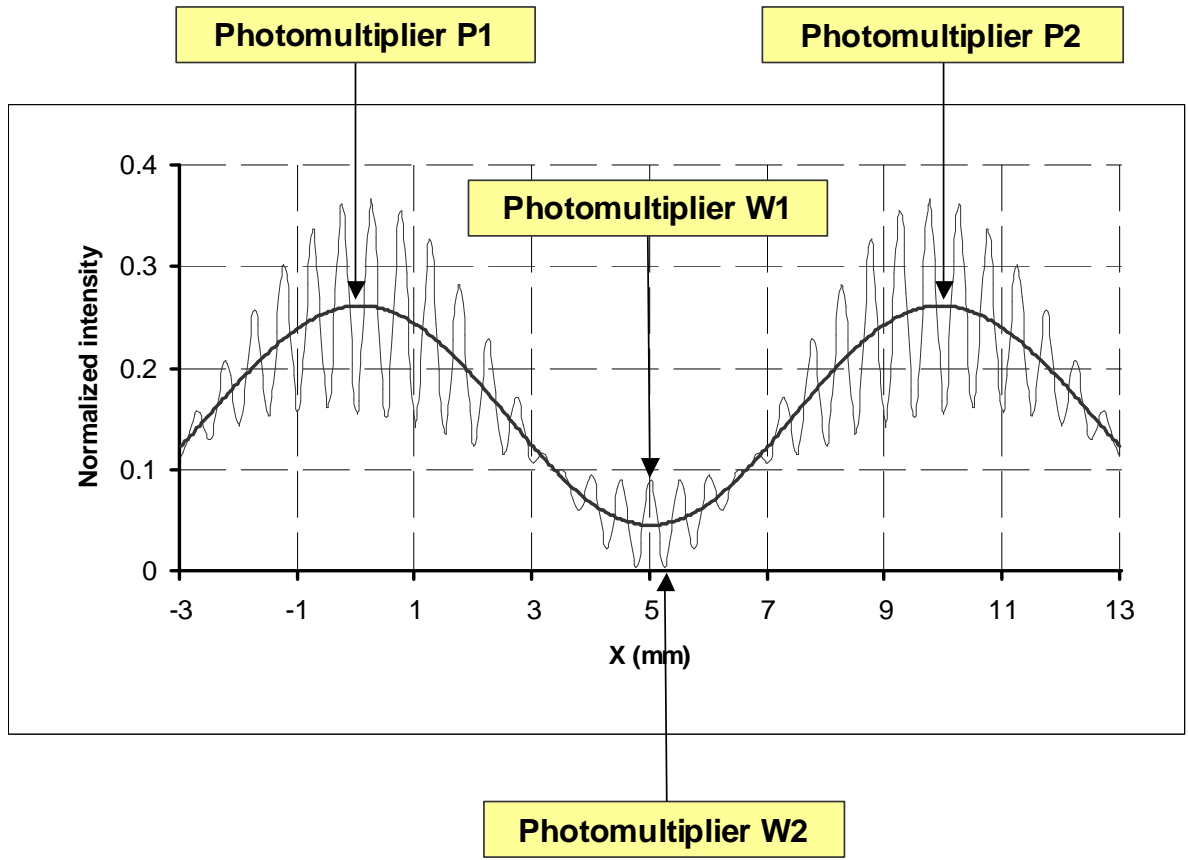

Figure 3: Locations of the four optical fiber heads in the observation plane with respect to the intensity distributions revealing an interference pattern (thin line) or not (thick line). Intensity curves are normalized so that their maxima are equal to 1 when both beams are perfectly superimposed.

According to common logic, the possible outcomes of the proposed *Gedanken* experiment are exhaustively summarized in Table 1, any historical theory being not excluded *a priori* (i.e. light can be described as a particle only, a wave only, either a wave or a particle, or both at the same time). We note that in the single-photon regime, the simplified model defined in section 2 would most probably imply a corpuscular-like behavior (second line of Table 1). This would also lead to consider the principle of complementarity in a different manner: if interference can only originate from separate light quanta following different paths, then the which-way information P is no longer relevant, and in that sense the principle of complementarity would become an evidence.

Whatever could be the actual results of such an experiment if it is conducted someday, it must be highlighted that the proposed setup would also be helpful in view of confirming the theoretical inequality $P^2 + W^2 \leq 1$, which is the modern form of the principle of complementarity. It could also be operated in delayed choice mode (for sufficiently long distances separating the entrance prism from mirror M), or considered as an alternative version of quantum eraser experiments [16], that might be easier to implement. Finally, the proposed measurement setup could easily be switched onto a pure beam splitting coincidence experiment [4] by means of a chopping or another equivalent fast blocking device located in the separation plane noted S in Figure 2.

| Basic principle | Expected result for P | Expected result for W | Scientific School |
|---|---|---|---|
| Photon is a pure particle. Interference does not occur | P is always equal to 1 | W is always equal to 0 | Newtonian |
| Photon is made of light quanta only interfering together | P is always equal to 1 | W is always equal to 0 | |
| Photon is a pure wave that can be divided and cannot be localized | P is always equal to 0 | W is always equal to 1 | Maxwell's equations |
| Photon is either a particle or a wave, but never at the same time | $P^2 + W^2 < 1$ | | Bohr's principle of complementarity |
| Photon is both a particle and a wave at the same time | P is always equal to 1 | W is always equal to 1 | De Broglie and Bohm's "pilot wave" |

Table 1: Possible outcomes of the *Gedanken* experiment in single-photon regime.

## 4  CONCLUSION

Questions about the nature of light are haunting humanity since thousand of years, as demonstrated by an old Egyptian representation on the funeral stele of Lady Taperet (9th-10th century BC) reproduced in Figure 4. Here the Lady stands in front of the Sun god Ra-Horakhty illuminating her with a bundle of rays made of flowers. These flowers could either be interpreted as single particles (i.e. photons), or as light quanta behaving together as a wave.

Here we should not preclude any of the possible outcomes of the proposed *Gedanken* experiment operated in single or starved photon regime, even if a particle-like behavior would be in agreement with the herein described model where photons and light quanta are considered as different objects. For now, reversing Dirac's famous statement is just a hypothesis that could clarify some apparent paradoxes, partly originating from our human senses. As German philosopher Immanuel Kant wrote in his *Critique of Pure Reason*: "What may be the nature of objects considered as things in themselves […] is quite unknown to us. We know nothing more than our own mode of perceiving them, which is peculiar to us." But it is fair to let the final words to P. A. M. Dirac himself, writing "the main object of physical science is not the provision of pictures, but is the formulation of laws governing phenomena and the application of these laws to the discovery of new phenomena" [5]. Perhaps new experiments such as the one presented in this paper may help to bring the picture of "light quanta only interfering with others" into real physical science.

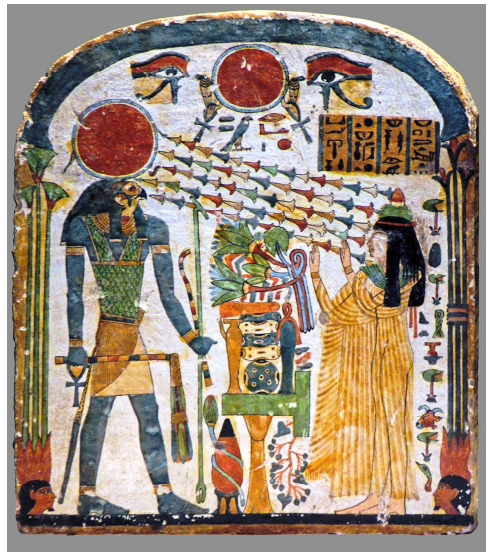

Figure 4: Stele of Lady Taperet, 9th-10th century BC (Credit Musée du Louvre/C. Décamps).